\documentclass[aps,prl,twocolumn,superscriptaddress,floatfix]{revtex4-1}

\usepackage{physics}
\usepackage{graphicx}
\usepackage{booktabs}
\usepackage{bbm}
\usepackage{bm}
\usepackage{siunitx}
\renewcommand{\figurename}{{\bf Fig.}}

\newcommand{\bscco}{Bi$_2$Sr$_2$CaCu$_2$O$_8$ }
\newcounter{para}
\newcommand\mypara{ \par\refstepcounter{para}\noindent} 
\renewcommand\vec{\mathbf}
\newcommand\QP{$\vec{Q}_{P}$ }
\newcommand\DPQ{$\Delta_P^{\vec{Q}}$ }

\usepackage{hyperref}
\hypersetup{
    colorlinks=true,
    linkcolor=blue,     
    urlcolor=blue,
    citecolor=blue
}
\begin{document}

\title{Magnetic-field Induced Pair Density Wave State 
in the Cuprate Vortex Halo}

\author{S. D. Edkins}
\affiliation{LASSP, Department of Physics, Cornell University, Ithaca, NY 14853, USA} 
\affiliation{School of Physics and Astron., University of St. Andrews, Fife KY16 9SS, Scotland}
\affiliation{Department of Applied Physics, Stanford University, Stanford, CA 94305, USA}

\author{A. Kostin}
\affiliation{LASSP, Department of Physics, Cornell University, Ithaca, NY 14853, USA} 

\author{K. Fujita}
\affiliation{LASSP, Department of Physics, Cornell University, Ithaca, NY 14853, USA} 
\affiliation{CMPMS Department, Brookhaven National Lab., Upton NY, USA}

\author{A. P. Mackenzie}
\affiliation{School of Physics and Astron., University of St. Andrews, Fife KY16 9SS, Scotland}
\affiliation{Max-Planck Institute for Chemical Physics of Solids, D-01187 Dresden, Germany}

\author{H. Eisaki}
\affiliation{Inst. of Advanced Industrial Science and Tech., Tsukuba, Ibaraki 305-8568, Japan}

\author{S. Uchida}
\affiliation{Department of Physics, University of Tokyo, Bunkyo-ku, Tokyo 113-0033, Japan}

\author{Subir Sachdev}
\affiliation{Department of Physics, Harvard University, Cambridge, MA, 02138, USA}

\author{M. J. Lawler}
\affiliation{LASSP, Department of Physics, Cornell University, Ithaca, NY 14853, USA}
\affiliation{Department of Physics and Astronomy, Binghamton University, Binghamton, NY 13902, USA.}

\author{E. -A. Kim}
\affiliation{LASSP, Department of Physics, Cornell University, Ithaca, NY 14853, USA}

\author{J. C. S\'eamus Davis}
\affiliation{LASSP, Department of Physics, Cornell University, Ithaca, NY 14853, USA}
\affiliation{School of Physics and Astron., University of St. Andrews, Fife KY16 9SS, Scotland}
\affiliation{CMPMS Department, Brookhaven National Lab., Upton NY, USA}

\author{M. H. Hamidian}
\affiliation{LASSP, Department of Physics, Cornell University, Ithaca, NY 14853, USA}
\affiliation{Department of Physics, Harvard University, Cambridge, MA, 02138, USA}

\begin{abstract}

When very high magnetic fields suppress the superconductivity in underdoped cuprates, an exceptional new electronic phase appears. It supports remarkable and unexplained quantum oscillations and exhibits an unidentified density wave (DW) state. Although generally referred to as a ``charge'' density wave (CDW) because of the observed charge density modulations, theory indicates that this could actually be the far more elusive electron-pair density wave state (PDW). To search for evidence of a field-induced PDW in cuprates, we visualize the modulations in the density of electronic states $N(\vec{r})$ within the halo surrounding \bscco vortex cores. This reveals multiple signatures of a field-induced PDW, including two sets of $N(\vec{r})$ modulations occurring at wavevectors \QP  and $2\vec{Q}_P$, both having predominantly \emph{s}-symmetry form factors, the amplitude of the latter decaying twice as rapidly as the former, along with induced energy-gap modulations at $\vec{Q}_P$. Such a microscopic phenomenology is in detailed agreement with theory for a field-induced primary PDW that generates secondary CDWs within the vortex halo. These data indicate that the fundamental state generated by increasing magnetic fields from the underdoped cuprate superconducting phase is actually a PDW with approximately eight CuO$_2$ unit-cell periodicity ($\lambda=8a_0$) and predominantly \emph{d}-symmetry form factor.

\end{abstract}

\maketitle

\mypara  
Cooper-pairs, if they have finite center-of-mass momentum $\vec{Q}_P$, can form a remarkable state in which the density of pairs modulates periodically in space at wavevector \QP\cite{Fulde1964,Larkin1964}. Intense theoretical interest has recently emerged \cite{Berg2007,Lee2014a,Agterberg2015} in whether such a ``pair density wave'' (PDW) state could, due to strong local electron-electron interactions \cite{Himeda2002,Raczkowski2007,Yang2009,Loder2010,Corboz2014,Cai}, be another principal state along with uniform superconductivity in the phase diagram of underdoped cuprates. This has been motivated by numerous experimental observations that can be understood in that context. For example, although intra-planar superconductivity appears in La$_{2-x}$Ba$_x$CuO$_4$ at relatively high temperatures, inter-planar superconductivity is strongly frustrated \cite{Li2007} in a fashion that could be explained by orthogonal unidirectional PDW states in each sequential CuO$_2$ plane \cite{Berg2007,Berg2009,Berg2009b}. Moreover, the measured momentum-space electronic structure of the cuprate pseudogap phase is consistent with predictions based on a biaxial PDW \cite{Lee2014a}. Reported breaking of time-reversal symmetry could be due to a PDW with inversion breaking, either alone or entangled with a CDW \cite{Pepin2014,Freire2015c,Wang2015b,Agterberg2015b}. The field-induced momentum-space reconstruction and consequent quantum oscillation phenomenology may be analyzed in terms of a PDW state \cite{Zelli2011,Zelli2012}. At highest fields, strong diamagnetism in torque magnetometry \cite{Yu2016} and supercurrents in DC transport  \cite{Yu-TeHsu2017} might both be understood as due to a field-induced PDW state. Most recently, scanned Josephson tunneling microscopy allows direct visualization of cuprate PDW modulations \cite{Hamidian2016}. Taken together, these studies indicate that a PDW may exist in underdoped cuprates, with the most common model invoked being an eight unit-cell (8$a_0$) periodic modulation of the electron-pair condensate. \\
\mypara
Such a PDW state clearly does not predominate at low temperature in zero magnetic field where global \emph{d}-wave superconductivity is robust. However, application of high magnetic fields appears to destabilize cuprate superconductivity towards an unidentified DW state \cite{Wu2011,Chang2012a,LeBoeuf2012,Canosa2013,Wu2013,Gerber2015,Chang2016,Jang2016} coincident with unexplained quantum oscillations \cite{Vignolle2013,Sebastian}. Among the peculiar characteristics of this DW are that, while magnetic fields amplify its intensity dramatically, this only occurs when superconductivity is present \cite{Wu2011,Chang2012a,LeBoeuf2012,Canosa2013,Wu2013,Gerber2015,Chang2016,Jang2016}; and that it emerges at highest fields as a distinct but unidentified long-range ordered state \cite{Wu2013,Gerber2015,Chang2016,Jang2016}. For type-II superconductors in general, application of magnetic field generates quantized vortices. Moreover, translational symmetry breaking is known to occur in the ``halo'' region surrounding the cuprate vortex core \cite{Hoffman2002a,Matsuba2007,Yoshizawa,Machida2016}, leading to a variety of hypotheses for the identity of the primary field-induced DW state therein \cite{Agterberg2015,Kivelson2002,Agterberg2009,Sau,Einekel,Wang2018,Dai2018}. Among them is the possibility that this state is not a conventional CDW but, instead, could be a PDW \cite{Lee2014a,Agterberg2015,Agterberg2009,Wang2018,Dai2018}. This is a fundamental distinction because the PDW and CDW are extremely different states in terms of broken symmetries and many-body wavefunctions, and because the wavevector of observed charge modulations \cite{Wu2011,Chang2012a,LeBoeuf2012,Canosa2013,Wu2013,Gerber2015,Chang2016,Jang2016,Vignolle2013} could actually be 2\QP, where \QP is the true wavevector of the pair density wave state. Thus, to determine whether the primary field-induced state of cuprates is a PDW has recently become an urgent research challenge. \\
\mypara
To search for evidence of such a state, we study the field-induced modulations of the density of electronic states $N(\vec{r},E)$ within the halo surrounding quantized vortex cores \cite{Hoffman2002a,Matsuba2007,Yoshizawa,Machida2016}. Any periodic modulations of electronic structure can be described by $A(\vec{r})=AF(\theta)  \cos(\vec{Q}\cdot\vec{r}+\phi_0)$, where $A(\vec{r})$ represents the modulating electronic degree of freedom with amplitude $A$, $\vec{Q}$ is the wavevector, and $F(\theta)$ is the modulation form factor. Of relevance here, is that an \emph{s}-symmetry form factor $F_s(\theta)$ is even under 90$^{\circ}$ rotations whereas a \emph{d}-wave form factor $F_d (\theta)$ is odd. The order parameters we consider are those of homogenous \emph{d}-wave superconductivity $\Delta(\vec{r})=F_{SC}\Delta_{SC}$  with $F_{SC}=F_d$, and that of a pair density wave $\Delta_{PD}(\vec{r})=F_P\Delta^{\vec{Q}}_{P}[e^{i\vec{Q}_P\cdot\vec{r}}+e^{-i\vec{Q}_P\cdot\vec{r}}]$ with wavevector \QP and either type of form factor (see Materials \& Methods Section 1, Ref. \cite{MM}). A field-induced PDW may be identified based on Ginzburg-Landau (GL) analysis \cite{Agterberg2015} of the interactions between these two OP within the halos of suppressed but non-zero superconductivity, that surrounds vortex cores. Given a generic GL free energy density of the form
\begin{equation}
\mathcal{F}_{\textrm{PDW-dSC}}=\mathcal{F}_{\Delta_{SC}}+\mathcal{F}_{\Delta_{A}}+u_{1}|\Delta_{A}|^{2}|\Delta_{SC}|^{2}|
\end{equation}\\
where  $F_{\Delta_{SC}}$ and $F_{\delta_{A}}$ are the free energy densities of a superconductor and of an alternative repulsively-coupled ($u_1>0$) state $\Delta_{A}$, observation of coexistence of $\Delta_A$ with $\Delta_{SC}$ within the vortex halo \cite{Hoffman2002a,Matsuba2007,Yoshizawa,Machida2016} already contains important information (Materials \& Methods Section 2, Ref. \cite{MM}). This is because the second state can only be stabilized in the halo region if the two ordered states are almost energetically degenerate \cite{Kivelson2002}. Such a near degeneracy occurs most naturally between a superconductor $\Delta_{SC}$ and a PDW \DPQ that are made up of the same electron-pairs.  In this case, $N(\vec{r})$ modulations generated by interactions between them can be found from products of these order parameters that transform as density-like quantities. The first of these is the product of PDW and uniform SC order parameters
\begin{equation}
    A_{\vec{Q}_P} \propto \Delta^{\vec{Q}}_{P}\Delta_{SC}^* \Rightarrow N(\vec{r})\propto \cos(\vec{Q}\cdot\vec{r})
\end{equation}
resulting in $N(\vec{r})$ modulations at the PDW wavevector \QP. Thus, a PDW induced in a vortex halo \cite{Agterberg2015,Wang2018,Dai2018} should produce the effects represented by Eqn. 2 and shown schematically in Fig. 1A.  Of key relevance to this study is the product of a robust PDW with itself:
\begin{equation}
    A_{2\vec{Q}_P} \propto \Delta_{P}^{\vec{Q}}\Delta_{P}^{-\vec{Q}*} \Rightarrow N(\vec{r}) \propto \cos(2\vec{Q}\cdot\vec{r})
\end{equation}
because this combination produces $N(\vec{r})$ modulations occurring at 2\QP. Moreover, one would expect the $N(\vec{r})$ modulations at \QP and 2\QP to coexist if a PDW is induced in a vortex halo, as shown schematically in Fig. 1B. Thus, a key signature of a field-induced PDW would be the appearance in vortex halos of $N(\vec{r})$, and thus charge density, modulations occurring at $\vec{Q}_P$ and simultaneously at 2$\vec{Q}_P$. \\
\mypara
In theory, significant further information can be determined from measured rates of decay of the induced $N(\vec{r})$ modulations away from the vortex center, and from the form factors of these modulations within the vortex halo. For a field-induced PDW, the $N(\vec{r},E)$ modulations at 2\QP should decay at twice the rate as those at \QP. This is because, if    $\Delta_P^{\vec{Q}}=\Delta_P^{\vec{Q}}(|\vec{r}|=0)e^{-|\vec{r}|/\xi}$, then $\Delta_P^{\vec{Q}}\Delta_P^{-\vec{Q}*}$ decays with $|\vec{r}|$ at twice the rate of $\Delta_P^{\vec{Q}}\Delta_{SC}^*$, as shown schematically in Fig. 1B. More importantly, while the $N(\vec{r},E)$ modulations at 2\QP due to $\Delta_P^{\vec{Q}}\Delta_P^{-\vec{Q}*}$ are always of $F_s$  form factor because they are the product of two identical order parameters,  if the  $N(\vec{r},E)$ modulations at \QP due to $\Delta_P^{\vec{Q}}\Delta_{SC}^*$ are predominantly of \emph{s}-symmetry form factor ($F_s$), this would reveal that the PDW order parameter $\Delta_P^{\vec{Q}}$ has a \emph{d}-symmetry form factor ($F_d$), and vice versa (see below and Material \& Methods Section 3, Ref. \cite{MM}). Overall then, since microscopic theory predominantly predicts a \emph{d}-symmetry form factor PDW for cuprates \cite{Himeda2002,Raczkowski2007,Yang2009,Loder2010,Corboz2014,Cai}, its signature in a vortex halo should be two sets of $N(\vec{r})$ modulations occurring at $\vec{Q}_P$ and 2$\vec{Q}_P$, both having predominantly \emph{s}-symmetry form factors, and the amplitude of the latter decaying twice as rapidly as that of the former. \\
\mypara
To explore these predictions, we image scanning tunneling microscope (STM) tip-sample differential tunneling conductance $\frac{dI}{dV}(\vec{r},V)\equiv g(\vec{r},E)$, versus bias voltage $V=E/e$ and location $\vec{r}$ with sub-unit-cell spatial resolution. We follow the procedure of the classic Hoffman experiment \cite{Hoffman2002a} in which $N(\vec{r},E)$ is measured at zero field and then at high magnetic field $B$ in the identical field of view (FOV) using an identical STM tip. The former is subtracted from the latter to yield the field-induced changes $\delta g(\vec{r},E,B)$, which are related to the field-induced perturbation to the density of states as $\delta N(\vec{r},E,B)\propto \delta g(\vec{r},E,B)$.  In this study, the range of this technique is greatly extended by enhancing both the r-space resolution using smaller pixels and the $q$-space resolution by using larger FOV, by increasing the numbers of vortices, by using distortion-corrected sublattice-phase-resolved imaging \cite{Fujita2014a}, and by measuring in a far wider energy range. Specifically, the $g(\vec{r},E,B)$ are measured at $T=2$K for slightly underdoped \bscco samples ($T_c\approx88$K; $p\approx17$\%)) and for $0<|E|<80$meV in magnetic fields up to $B=8.25$T in a 65nmx65nm field-of-view. The $g(\vec{r},E,B)$ data are acquired in precisely the same FOV using an identical STM tip at $B=0$ and $B=8.25$T. Then, each image is distortion corrected \cite{Fujita2014a} to render the atomic lattice perfectly periodic, and finally registered to each other within every CuO$_2$ unit cell with $\sim$30 pm precision (see Materials \& Methods Section 4, Ref. \cite{MM}). The two resulting $g(\vec{r},E,B)$ data sets are subtracted to yield the field-induced effects on electronic structure in $\delta g(\vec{r},E,B)=g(\vec{r},E,B)-g(\vec{r},E,0)$. This final key step results in studying phenomena that are uniquely those induced by magnetic fields \cite{Hoffman2002a}, and with the signatures of the ubiquitous \emph{d}-symmetry form factor DW that occurs in all samples at $B=0$ having been subtracted.\\
\mypara
The location of every vortex halo in $\delta g(\vec{r},E,B)$ images is next identified by using two well-known phenomena: (i) suppression of the superconducting coherence peaks (Fig. 2B) and, (ii) appearance of periodic conductance modulations at $|E|<16\textrm{meV}$ \cite{Hoffman2002a,Matsuba2007,Yoshizawa,Machida2016}. Figure 2C shows measured $\delta g(\vec{r},10\textrm{meV})= g(\vec{r},10\textrm{meV},8.25\textrm{T})-g(\vec{r},10\textrm{meV},0\textrm{T})$ and illustrates excellent agreement with previous studies of low conductance modulations with $\vec{q}\approx(\pm \frac{1}{4},0);(0,\pm \frac{1}{4})\frac{2\pi}{a_0}$ in \bscco vortex halo \cite{Hoffman2002a,Matsuba2007,Yoshizawa,Machida2016}. In this study, we focus on a different energy range $25<|E|<50$meV because, as shown in Fig. 2B, the other major changes between a typical conductance spectrum at zero field (solid curve) and that at the center of a vortex at the same location (dashed curve), occur in this energy range (Materials \& Methods Section 5, Ref. \cite{MM}). In Fig. 3A we show measured $\delta g(\vec{r},30\textrm{mev})$ containing the modulations detected surrounding the centre point of each vortex core. Fourier analysis of this $\delta g(\vec{r},30\textrm{meV})$ yields $\left |\widetilde{\delta g}(\vec{q},30\textrm{meV})\right |$ as shown in Fig. 3B, with the immediate discovery of four sharp peaks at $\vec{q}=[\vec{Q}_{P}^{x};\vec{Q}_P^y]\approx[(\pm \frac{1}{8},0);(0,\pm \frac{1}{8})]2\pi/a_0$ which we label \QP for reasons explained below. Similarly, there is a second set of weaker modulations in $\widetilde{\delta g}(\vec{q},30\textrm{meV})$ at $\vec{q}\approx[(\pm \frac{1}{4},0);(0,\pm \frac{1}{4})]2\pi/a_0$ which we label 2\QP. The measured r-space amplitude-envelopes of the \QP and 2\QP modulations shown in Figs. 3C,D reveal how these field-induced phenomena are confined to the vortex halo regions only. Averaged over all vortices, the measured amplitude $\left |\widetilde{\delta g}(\vec{q},30\textrm{meV})\right |$ plotted along (1,0) in Fig. 3E discernibly discriminates the \QP from the 2\QP peaks. Thus, we discover strong field-induced modulations of $N(\vec{r},E)$ with period approximately 8$a_0$ coexisting with weaker modulations of period approximately 4$a_0$, along both the (1,0);(0,1) directions within every vortex halo. These phenomena exist within the energy range $25<|E|<45\textrm{meV}$. \\
\mypara 
To evaluate form factor symmetry for these field-induced modulations, we separate each such $\delta g(\vec{r},E)$ image into three sublattice images \cite{Fujita2014a}: $Cu(\vec{r},E)$, contains only the measured values of $\delta g(\vec{r},E)$ at copper sites and $O_{x}(\vec{r},E)$ and $O_{y}(\vec{r},E)$, contain only those at the x/y-axis planar oxygen sites. Here it is important to emphasize that all of these form factors refer to modulations in $\delta g(\vec{r},E,B)$ and are not necessarily those of the order parameter of the field-induced state that generates them. Complex-valued Fourier transforms of the $O_x(\vec{r},E)$ and $O_y(\vec{r},E)$ sublattice images, yield $\widetilde{O}_{x}(\vec{q},E)$; $\widetilde{O}_{y}(\vec{q},E)$. Then, modulations at any $\vec{Q}$ having \emph{d}-symmetry form factor $F_d$ generate a peak in $\widetilde{D}^{\delta g}(\vec{q},E)\equiv \widetilde{O}_{x}(\vec{q},E)-\widetilde{O}_{y}(\vec{q},E)$ at $\vec{Q}$, while those with \emph{s}-symmetry form factor $F_s$ generate a peak in $\widetilde{S}^{\delta g}(\vec{q},E)\equiv \widetilde{O}_{x}(\vec{q},E)+\widetilde{O}_{y}(\vec{q},E)+\widetilde{Cu}(\vec{q},E)$ at $\vec{Q}$. When the data in Figs. 3A,B are analyzed in this way using measured $\widetilde{S}^{\delta g}(\vec{q},30\textrm{meV})$, the field-induced $\delta g(\vec{r},E)$-modulations occurring at $\vec{Q}\approx (\pm Q_P,0);(0,\pm Q_P )$ and $Q\approx (\pm 2Q_P,0);(0,\pm 2Q_P)$ all exhibit \emph{s}-symmetry form factors. However, the measured $\widetilde{D}^{\delta g}(\vec{q},30\textrm{meV})$ in Figs. 4A,B also reveals that additional \emph{d}-symmetry $\delta g(\vec{r},E)$-modulations occur at $\vec{Q}\approx(0,\pm Q_P)$ and $\vec{Q}\approx(0,\pm2Q_P)$. They too are confined to the vortex halo as indicated by the r-space amplitude-envelope of the 2\QP-modulations in $\widetilde{D}^{\delta g}(\vec{q},30\textrm{meV})$ as shown in Fig. 4C.  \\
\mypara
Figures 5A,B show the overall measured amplitudes of $\left |\widetilde{\delta g}(\vec{q},30\textrm{meV})\right |$ derived from $\delta g(\vec{r},30\textrm{meV})$ in Fig. 3A, plotted along the (1,0) and (0,1) directions of the CuO$_2$ plane. Figures 5C,D show equivalent cuts of $\left |\widetilde{\delta g}(\vec{q},-30\textrm{meV})\right |$ derived from $\delta g(\vec{r},-30\textrm{meV})$ data. The four maxima at $|\vec{q}|\approx1/8$, $|\vec{q}|\approx1/4$, $|\vec{q}|\approx3/4$ and $|\vec{q}|\approx7/8$ due to the field induced modulations are evident. The measured form factor of each set of modulations is identified by color code, red being \emph{s}-symmetry and blue \emph{d}-symmetry. Although modulations at $|\vec{q}|\approx7/8$, $|\vec{q}|\approx3/4$ (blue Fig. 5A-D) appear subdominant, they do merit comment. First, they are not inconsistent with a small \emph{s}-symmetry component in the PDW order parameter as described by Eqns. 2 and 3 with form factor $F_s$. However, these phenomena may also represent a field-induced version of the unidirectional \emph{d}-symmetry form factor $N(\vec{r},E)$ modulation, as observed extensively in zero field \cite{Fujita2014a}. \\
\mypara
Nevertheless, the predominant phenomena detected are the two sets of \emph{s}-symmetry form factor modulations at $|\vec{Q}_P|\approx1/8$,  $|2\vec{Q}_P|\approx1/4$ (red Figs. 5A-D). The former \emph{s}-symmetry modulation is key because, when induced by $\Delta_P^{-\vec{Q}}\Delta_{SC}^{*}$, it is caused by a PDW order parameter $\Delta_P^{-\vec{Q}}$ which has \emph{d}-symmetry. Equally importantly, after subtraction of a smooth background, the widths $\delta \boldsymbol{q}$ of all $|\vec{Q}_P|\approx1/8$ peaks are about half of the $|2\vec{Q}_P|\approx1/4$ peaks, as determined quantitatively by fitting as shown in Figs. 5A-D. Averaged over the two directions (1,0) and (0,1) and energies $E=\pm30\textrm{meV}$, we find that $\delta(2\vec{Q}_P)=(1.8\pm0.2)\delta(\vec{Q}_P)$ consistent with a field-induced PDW (Fig. 1) \cite{Agterberg2015,Wang2018,Dai2018}. As an additional marker of a field induced PDW we consider whether in the locally-defined energy gap $\Delta(\vec{r})$ modulates as $\Delta(\vec{r})=\Delta_{SC}+\Delta_P\cos(\vec{Q}_P\cdot \vec{r})$. Empirically $\Delta(\vec{r})$ is defined by the energy of the peaks in $N(\vec{r},E)$ (horizontal arrow in Fig. 2A). In that case, the field-induced changes to the gap are defined as $\delta\Delta(\vec{r})= \Delta(\vec{r},8.25\textrm{T})-\Delta(\vec{r},0\textrm{T})$   (Materials \& Methods Section 6 Ref. \cite{MM}). When measured, $\delta\Delta(\vec{r})$ yields a Fourier transform $\widetilde{\delta\Delta}(\vec{q})$ as shown in Fig. 5E. This reveals a field-induced gap modulation at $\vec{Q}_P$ and not at 2\QP, as is expected specifically for a primary field-induced PDW at \QP. \\
\mypara
In sum, results shown in Figs. 3-5 indicate that, in \bscco, a field-induced pair density wave state emerges from the halo region surrounding each quantized vortex \cite{Agterberg2015, Wang2018, Dai2018}. The principal experimental signatures are two sets of $N(\vec{r})$ modulations occurring at \QP and 2\QP, both having \emph{s}-symmetry form factors, and the amplitude of the latter decaying twice as rapidly as that of the former. This inferred PDW has period very close to 8$a_0$,  is apparently bi-directional (see Materials \& Methods Section 7, Ref. \cite{MM}) and has \emph{d}-symmetry form factor. A range of important consequences stem from these observations. First and foremost, the primary state induced by high magnetic fields in superconducting cuprates is then a PDW with wavevector \QP, and it is accompanied by secondary charge modulations at \QP and 2\QP. Second, the 8$a_0$ periodicity points towards a strong-correlation driven microscopic mechanism for this PDW \cite{Himeda2002,Raczkowski2007,Yang2009,Loder2010,Corboz2014,Cai}. Third, because the PDW is enhanced by increasing magnetic field, our data imply that the high-field state of cuprates might itself be a PDW state \cite{Lee2014a} and, if so, it is likely phase fluctuating and intertwined with an additional CDW component. Finally, putting all such conjectures aside, we emphasize that the experimental observations in Figs. 3-5 are in excellent, detailed and quantitative agreement with theoretical models \cite{Agterberg2015,Wang2018,Dai2018} specifically of a field-induced primary PDW with \emph{d}-symmetry and wavevector \QP, that generates secondary CDWs at \QP and 2\QP, within the cuprate vortex halo. 
\newline \\
\textbf{Acknowledgements}

We acknowledge and thank D. Agterberg, P. Choubey, A. Chubukov, E. Fradkin, P.J. Hirschfeld, D.H. Lee, P.A. Lee, C. Pepin, S. Sebastian, S. Todadri, J. Tranquada and Yuxuan Wang, for helpful discussions, advice and communications. We are extremely grateful to S. A. Kivelson for crucial proposals on the complete set of PDW phenomena to search for within the vortex halo. J.C.S.D and M.H.H acknowledge support from the Moore Foundation’s EPiQS Initiative through Grant GBMF4544. SDE acknowledges studentship funding from the EPSRC under Grant EP/G03673X/1 and the Karel Urbanek postdoctoral fellowship at Stanford University; S.U. and H.E. acknowledge support from a Grant-in-Aid for Scientific Research from the Ministry of Science and Education (Japan); A.K., and K.F. acknowledge salary support from the U.S. Department of Energy, Office of Basic Energy Sciences, under contract number DEAC02-98CH10886. Experimental studies and analysis were carried out under the Moore Foundation’s EPiQS Initiative through Grant GBMF4544 and under contract number DEAC02-98CH10886 from the U.S. Department of Energy, Office of Basic Energy Sciences.
\newline \\
\textbf{Author Information}
Correspondence and requests for materials should be addressed to J.C.D. (jcseamusdavis@gmail.com) or M.H.H. (m.hamidian@gmail.com)

\bibliography{ref}

\begin{figure*}
\includegraphics[width=0.75\textwidth]{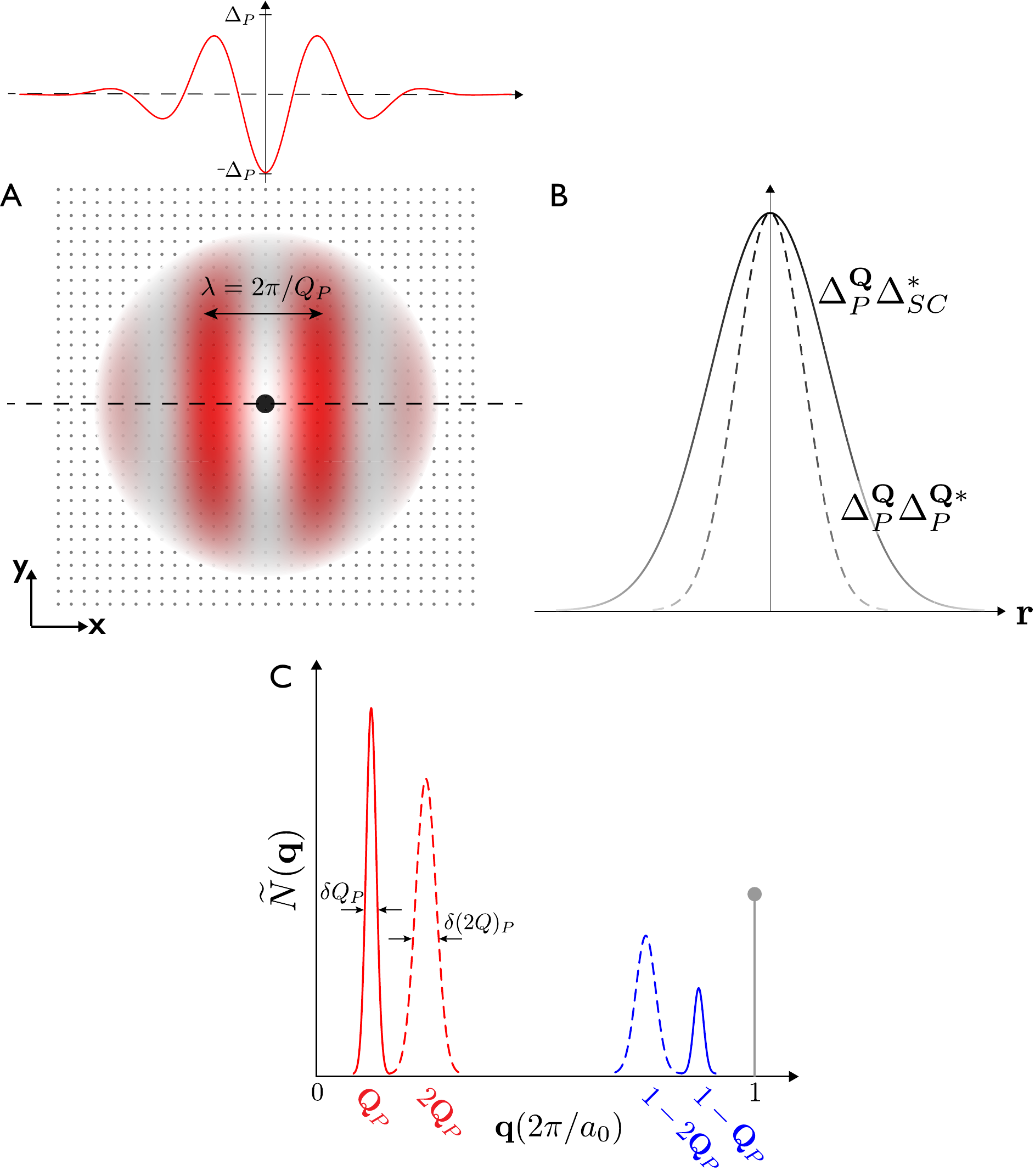}
\caption{\label{Fintro}{\bf Schematic of Field-induced Unidirectional $8a_0$ Pair Density Wave}
({\bf A}) Schematic diagram of the halo (grey) surrounding the vortex core where $\Delta_{SC}$ is completely suppressed (black) of a quantized vortex in a cuprate supeconductor (SC).  The CuO$_2$ plane orientiation and Cu-Cu periodicity is shown in the background. To analyze such a situation, within the halo we consider a PDW modulation along the x-axis with periodicity $8a_0$ as indicated schematically in red. The order parameter of homogeneous $d$-wave superconductivity $\Delta_{SC}$ rises quickly from zero near the symmetry point (black) to become non-zero in the halo region (and beyond).  The order parameter of the PDW is $\Delta_P^{\vec{Q}}(\vec{r})$ as in the text.
({\bf B}) The envelope containing non-zero amplitude $\Delta_P^{\vec{Q}}\Delta_{SC}^*$ of the induced $N(\vec{r})$ modulations due to the interaction between SC and PDW order parameters is shown by a solid curve, plotted along the dashed line in (A) through the vortex core.  The envelope containing non-zero amplitude $\Delta_P^{-\vec{Q}}\Delta_P^{\vec{Q}*}$ of the induced $N(\vec{r})$ modulations due to the PDW itself is shown by a dashed curve, plotted along the same dashed line in (A).  For pedagogical clarity, we ignore the small regions at the core where $\Delta_P^{\vec{Q}*}\Delta_{SC}$ must rise from zero as $\Delta_{SC}$ does; this is because the core radius in \bscco is only $\sim$1 nm.
({\bf C}) If the field-induced PDW has $d$-symmetry form factor, $F_P = F_d$, then two sets of \emph{s}-symmetry $N(\vec{r})$ modulations should appear together.  The first is $N(\vec{r}) \propto \cos(\vec{Q}_P\cdot \vec{r})$ due to $\Delta_P^{\vec{Q}} \Delta_S^*$ as indicated in $\widetilde{N}(\vec{q})$, the Fourier transform of $N(\vec{r})$, by a solid red curve.  The second $N(\vec{r}) \propto \cos(2\vec{Q}_P\cdot \vec{r})$ due to $\Delta_P^{\vec{-Q}}\Delta_P^{\vec{Q}*}$ is indicated in $\widetilde{N}(\vec{q})$  by a dashed red curve.  The decay length for the 2\QP modulation should be half that of the \QP modulation, meaning that the linewidth $\delta(2Q)_P$ of the 2\QP modulation (dashed red) should be twice that of the \QP modulation, $\delta Q_P$ (solid red).  If the PDW has $s$-symmetry form factor, $F_P=F_s$, then a different pair of $N(\vec{r})$ modulations should appear together. First is $N(\vec{r}) \propto \cos((\vec{Q}_B - \vec{Q}_P) \cdot \vec{r})$ due to $\Delta_P^{-\vec{Q}}\Delta_{SC}^*$ (solid blue line) and second $N(\vec{r}) \propto \cos(2\vec{Q}_P\cdot \vec{r})$ due to $\Delta_P^{-\vec{Q}}\Delta_P^{\vec{Q}*}$.}
\end{figure*}
\begin{figure*}
\includegraphics[width=0.95\textwidth]{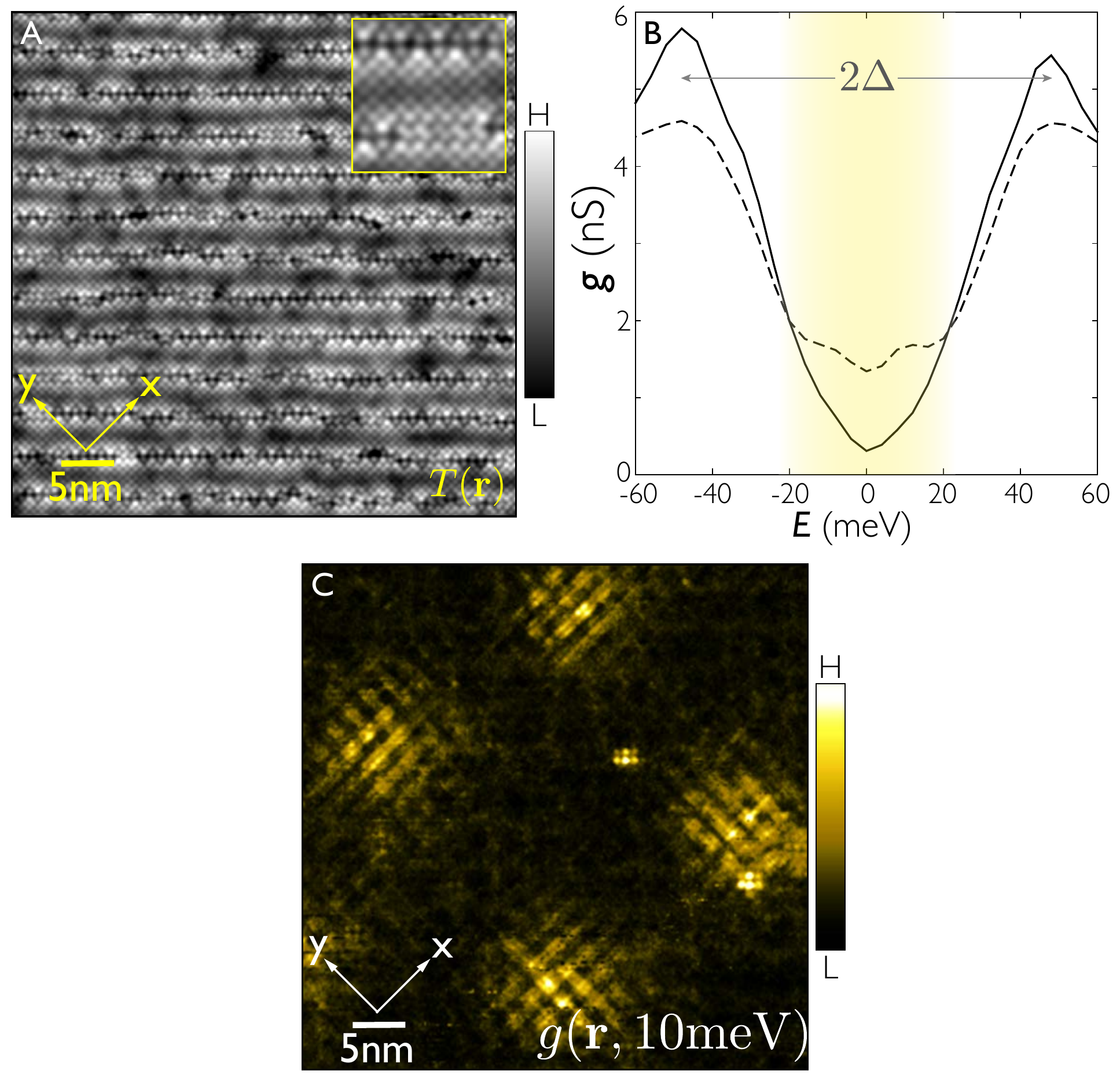}
\caption{\label{Fintro}{\bf Four-unit-cell Quasiparticle Modulation at Vortex Halos in \bscco}
({\bf A}) Topographic image $T(\vec{r})$ of BiO termination layer of the \bscco sample used in these studies.  It contains the locations of $\sim$6000 individually identified Cu sites and $\sim$12000 individually identified O sites within the CuO$_2$ plane beneath this surface.  The displacement of every specific atomic site in this field of view between zero field and $B=8.25$ Tesla was constrained by post processing of all low/high field data sets to be $\sim$ 30 pm. 
({\bf B}) Measured differential tunneling conductance spectrum $g(\vec{r},E = eV) \equiv dI/dV(\vec{r},V)$ at the symmetry point of a vortex core (dashed line) and at the identical location in zero field. There are two energy ranges where the introduction of the vortex impacts $N(\vec{r},E)$, at low energy (yellow) where Bogoliubov quasiparticle modulations are well known \cite{Hoffman2002a,Matsuba2007,Yoshizawa,Machida2016} and near the gap edge 25$<E<$50 meV which is the energy range studied here.
({\bf C}) Measured $\delta g(\vec{r},12 \mbox{meV}) = g(\vec{r},12\mbox{meV}, B = 8.5\mbox{T}) - g(\mathbf{r},12\mbox{meV},B=0\textrm{T})$ showing the four-unit-cell periodic pattern of quasiparticle states surrounding vortex cores in \bscco\cite{Hoffman2002a,Matsuba2007,Yoshizawa,Machida2016}.  Vortices are easily located thus, as with the four shown clearly in this FOV.}
\end{figure*}
\begin{figure*}
\includegraphics[width=0.95\textwidth]{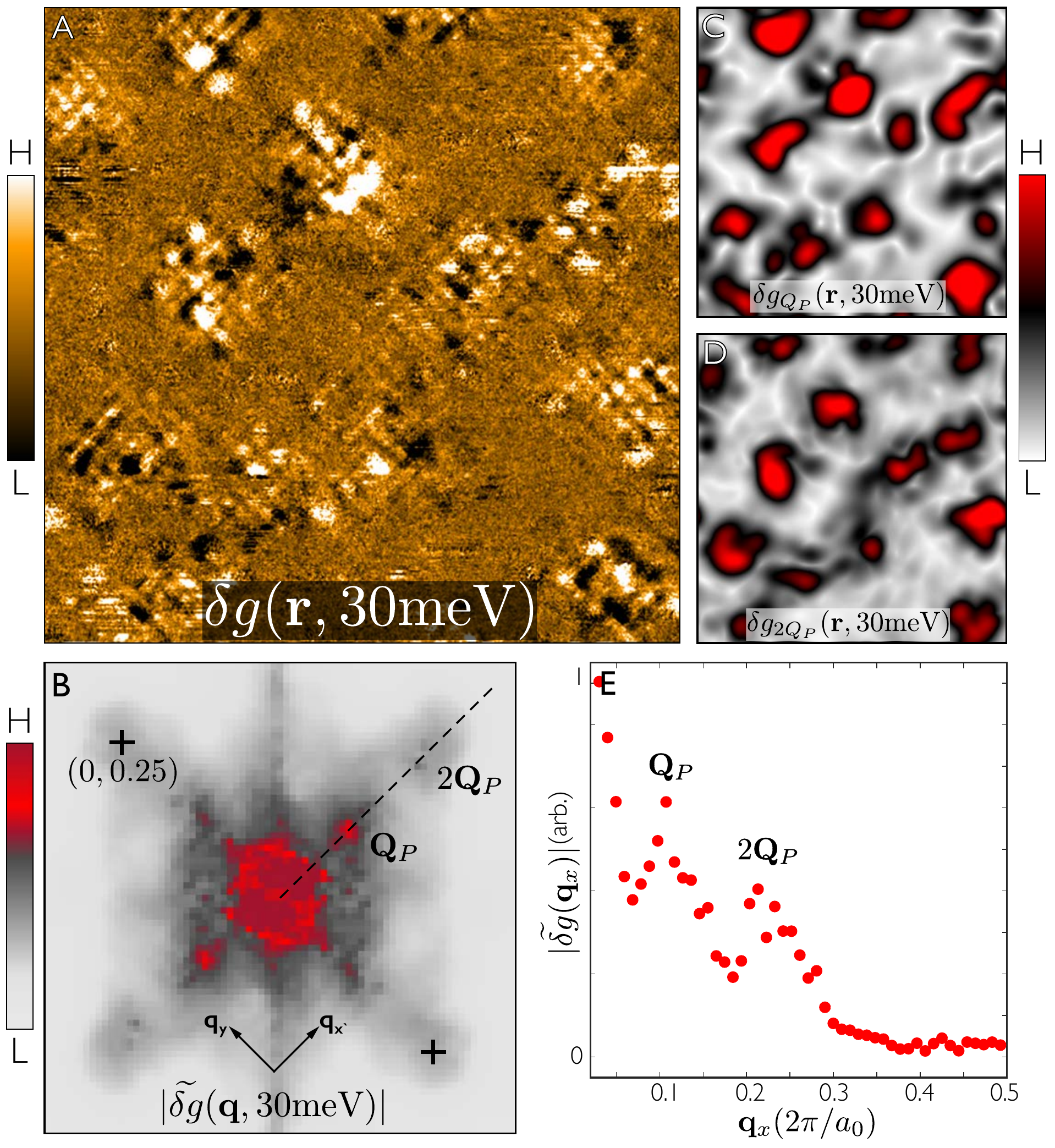}
\caption{\label{Fintro}{\bf Field induced \emph{s}-symmetry Form Factor Modulations Within Vortex Halos}
({\bf A}) Measured field-induced modulations $\delta g(\vec{r},30\mbox{meV},B = 8.25\mbox{T}) - g(\vec{r},30\mbox{meV},B = 0\mbox{T})$ in a 58nm$\times$58nm FOV. The simultaneously measured topographs $T(\vec{r})$ at $B=8.25$ T and 0T are show in Materials and Methods 3 \cite{MM}.
({\bf B}) Amplitude Fourier transform $|\widetilde{\delta g}(\vec{q},30 \mbox{meV})|$ (square root of power spectral density) of $\delta g(\vec{r},30\mbox{meV})$ in (A).  The $\vec{q} = [(0,\pm 1/4)]2\pi/a_0$ points are indicated by black crosses.  Four sharp maxima, indicated by \QP, occur at $\vec{q} \approx [(\pm 1/8,0);(0,\pm 1/8)]2\pi/ a_0$ while four broader maxima, indicated by 2\QP, occur at $\vec{q} \approx [(\pm 1/4,0);(0,\pm 1/4)]2\pi/ a_0$.
({\bf C}) Measured amplitude envelope of the modulations in $\delta g(\vec{r},30\mbox{meV})$ at \QP showing that they only occur within the vortex halo regions.
({\bf D}) Measured amplitude envelope of the modulations in $\delta g(\vec{r},30\mbox{meV})$ at 2\QP showing that they also only occur within the vortex halo region.
({\bf E}) Measured $|\widetilde{\delta g}(\vec{q},30\mbox{meV})|$ along $(0,0)-(1/2,0)$ (dashed line in (B)) showing the two maxima in the field induced $N(\vec{q})$ modulations occurring at \QP $=0.117\pm0.01$ and 2\QP $=0.231\pm 0.01$ (see Figs. 5A-D).
}
\end{figure*}
\begin{figure*}
\includegraphics[width=0.95\textwidth]{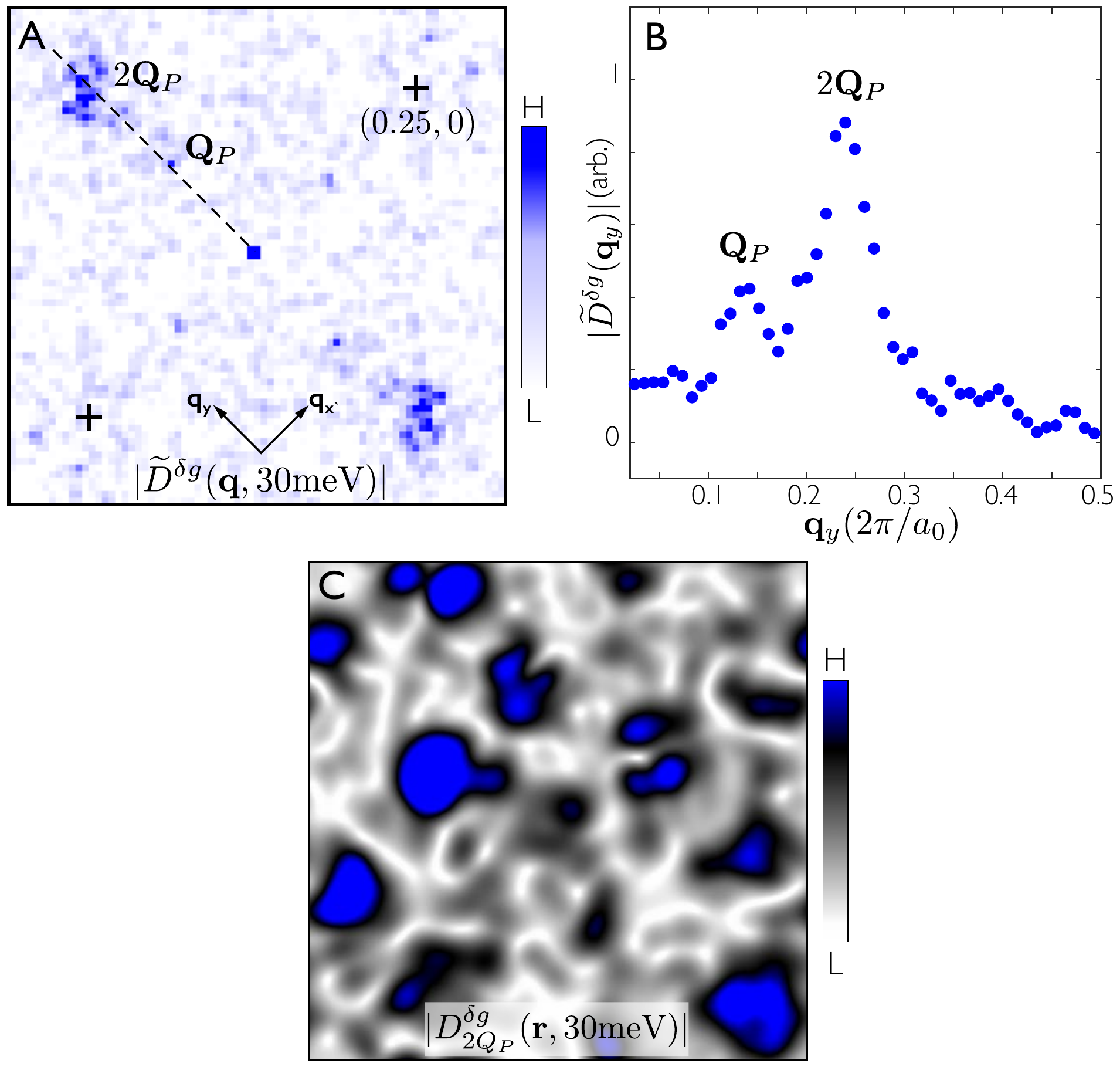}
\caption{\label{Fintro}{\bf Subdominant \emph{d}-symmetry Form Factor Modulations within Vortex Halos}
({\bf A}) Amplitude Fourier transform of the $d$-symmetry form factor modulations in $N(\vec{r}), |\widetilde{D}^{\delta g}(\vec{q},30\mbox{meV})|$, derived from measured $\delta g (\vec{r},30\mbox{meV})$ data in Fig. 3A. Again, $\vec{q} = [(0,\pm1/4)]2\pi/ a_0$ points are indicated by black crosses.  Two sharp maxima, indicated by $\vec{Q}_P$, occur at $\vec{q} \approx [(\pm 1/8,0);(0,\pm 1/8)]2\pi/ a_0$ while two broader maxima, indicated by $2\vec{Q}_P$, occur at $\vec{q} \approx [(0,\pm 1/4)]2\pi/ a_0$, both sets oriented along the y-axis.
({\bf B}) Measured $|\widetilde{D}^{\delta g} (\vec{q},30\mbox{meV})|$ along $(0,0)-(1/2,0)$ (dashed line in (A)) showing the maxima in the field induced $N(\vec{r})$ modulations occuring at \QP and 2\QP.  A unidirectional $d$-symmetry form factor charge density modulation, as observed extensively in zero field\cite{Fujita2014a}, would have such characteristics, as would an $s$-symmetry form factor PDW.  These modulaton did not appear in Fig. 3 because, in that unprocessed $\delta g(\vec{r},E)$ data, they occur at $\vec{Q} \approx (0,\pm 7/8)2\pi /a_0$ and $\vec{Q} \approx (0,\pm 3/4)2\pi/ a_0$ due to their $d$-symmetry form factor\cite{Fujita2014a}; see Figs. 5A-D.
({\bf C}) Measured amplitude envelope of the modulations in $|\widetilde{D}^{\delta g}(\vec{q},30\mbox{meV})|$ at 2\QP showing that these phenomena also only occur within the vortex halo regions as in Figs. 3C,D.
}
\end{figure*}
\begin{figure*}
\includegraphics[width=0.95\textwidth]{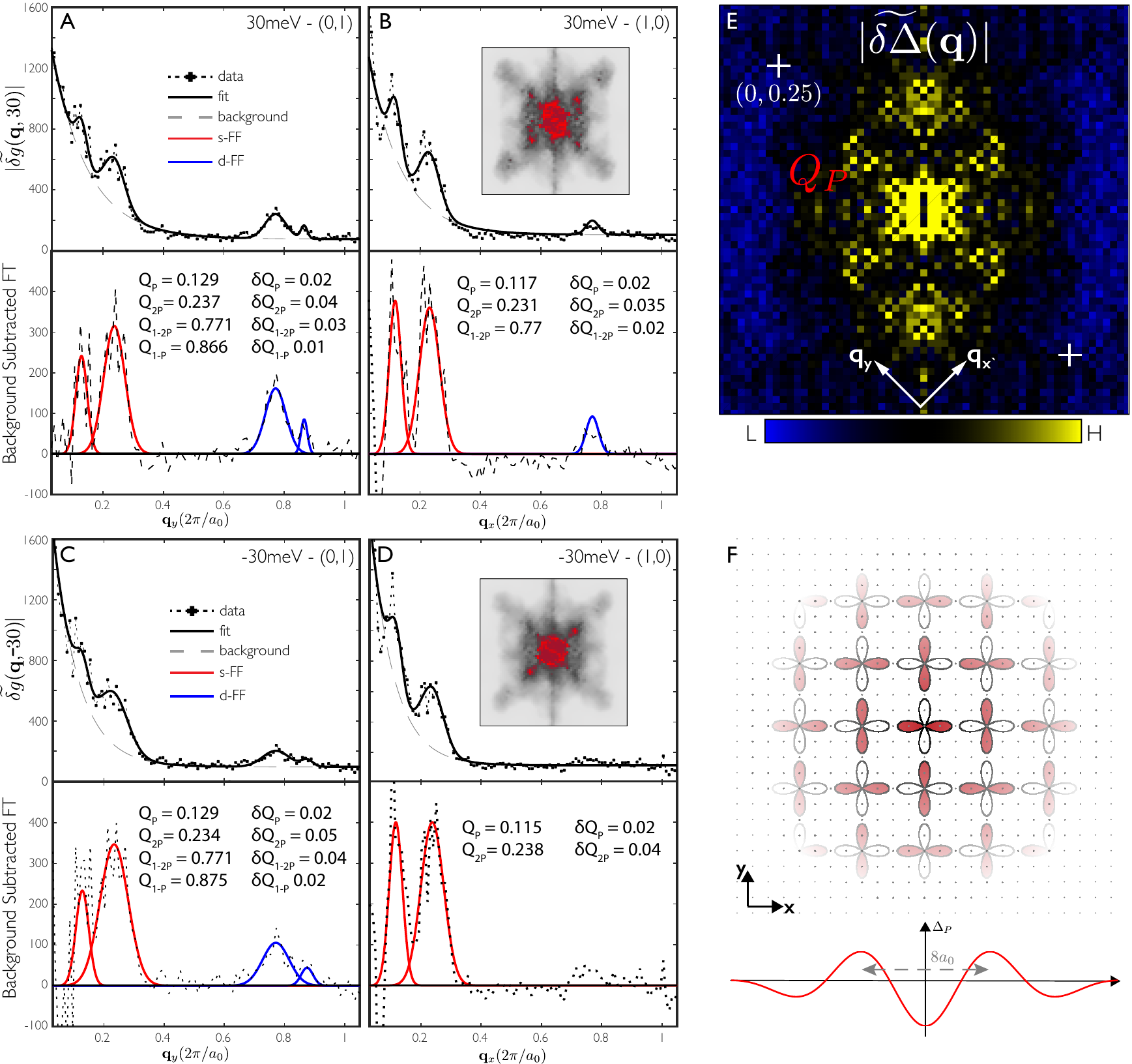}
\caption{\label{Fintro}{\bf Field-induced $\boldsymbol{N(\vec{r})}$ Modulations Indicate Pair Density Wave in Vortex Halo}
({\bf A,B}) Amplitude Fourier transform $|\widetilde{\delta g}(\vec{q},30\mbox{meV})|$, derived from $\delta g (\vec{r},30 \mbox{meV}$, data is plotted along two orthogonal axes from $(0,0)$-$(0,1)$ and $(0,0)$-$(1,0)$, to reach both Bragg points.  All four local maxima, \QP and 2\QP from the $s$-symmetry field induced $N(\vec{r})$ modulations, plus \textbf{1}-\QP and \textbf{1}-2\QP from the $d$-symmetry field induced $N(\vec{r})$ modulations are seen. Measurements from these fits of the $q$-magnitude and widths $\delta q$ of the $s$-symmetry peaks at \QP and 2\QP yields: $Q_P^x = 0.117, Q_P^y = 0.129;\ 2Q_P^x = 0.231, 2Q_P^y = 0.237;\ \delta Q_P^x = 0.020, \delta Q_P^y = 0.020;\ \delta (2Q_P^x) = 0.034, \delta(2Q_P^y) = 0.035$.  Inset shows $|\widetilde{\delta g}(\vec{q},30\mbox{meV})|$.
({\bf C,D}) Amplitude Fourier transform $|\widetilde{\delta g}(\vec{q},-30\mbox{meV})|$, derived from $\delta g (\vec{r},-30 \mbox(meV)$, again shows four local maxima at \QP and 2\QP from $s$-symmetry field induced $N(\vec{r})$ modulations, plus \textbf{1}-\QP and \textbf{1}-2\QP from the $d$-symmetry field induced $N(\vec{r})$ modulations.  Measurement yields: $Q_P^x = 0.115, Q_P^y = 0.128;\ 2Q_P^x = 0.239, 2Q_P^y = 0.235;\ \delta Q_P^x = 0.020, \delta Q_P^y = 0.020; \delta (2Q_P^x) = 0.039, \delta(2Q_P^y) = 0.045$.  Inset shows $|\widetilde{\delta g}(\vec{q},-30\mbox{meV})|$.  The $s$-symmetry field induced $N(\vec{r})$ modulations at \QP and 2\QP are almost particle particle-hole symmetric (insets (B),(D)) in the sense that $N(\vec{r},E>25\mbox{meV}) = N(\vec{r},E < -25\mbox{meV})$ for these two wavevectors.  This situation is diametrically opposite to the phenomenology of the $d$-symmetry form factor DW observed at zero field \cite{Fujita2014a}.
({\bf E}) Fourier transform $\widetilde{\delta \Delta}(\vec{q})$ of measured $\delta \Delta (\vec{r}) = \Delta(\vec{r},8.25\mbox{T}) - \Delta(\vec{r},0)$ (Materials and Methods Section 5, \cite{MM}). The observed peaks revealing field induced gap modulation occur at points indistinguishable from \QP. The peak along $(1,1)$ direction is at the wavevector of the crystal supermodulation, where the gap modulations due to unit cell geometry changes are well known. 
({\bf F}) Schematic representation of a bi-directional PDW with $d$-symmetry form factor induced within a vortex halo that is consonant with the data in this paper.}
\end{figure*}
\clearpage
\newpage
\onecolumngrid
\begin{center}
\vspace{2cm}
\textbf{\large Supplementary Materials}
\vspace{0.5cm}
\end{center}
\twocolumngrid
\setcounter{equation}{0}
\setcounter{figure}{0}
\setcounter{table}{0}
\setcounter{page}{1}
\makeatletter
\renewcommand{\theequation}{S\arabic{equation}}
\renewcommand{\figurename}{{\bf Figure S}}
\renewcommand{\tablename}{{\bf Table S}}

\section{Materials and Methods}
For our studies, high-quality Bi$_2$SrCaCu$_2$O$_{8+\delta}$ single crystals were grown using the travelling-solvent-floating zone (TSFZ) method. The samples are of Bi$_{2.1}$Sr$_{1.9}$CaCu$_2$O$_{8+\delta}$ and were synthesized from dried powders of Bi$_2$O$_3$, SrCO$_3$, CaCO$_3$ and CuO. The crystal growth was carried out in air and at growth speeds of 0.15-0.2 mm/h for all samples. Inductively coupled plasma (ICP) spectroscopy was used for the composition analysis and a vibrating sample magnetometer (VSM) was used for measurement of $T_c$. 
Here we studied samples of Bi$_2$SrCaCu$_2$O$_{8+\delta}$ with hole doping p $\approx$ 0.17. Each sample was inserted into the cryogenic ultra high vacuum of the SI-STM system and cleaved to reveal an atomically flat BiO surface. All measurements were performed at a temperature of 2K. The basic spectroscopic imaging STM consists of lock-in amplifier measurements of the differential tunneling conductance with sub-unit-cell resolution and register, as a function of both location $\vec{r}$ and electron energy $E$. We vary the applied magnetic field perpendicular to the CuO$_2$ planes of the samples using a superconducting solenoid with a highly stable persistent current/field. 

\section{Supplementary Text}
 
\subsection{1. Order Parameter Description of Pair Density Waves}

Phenomenologically we can describe pair density wave (PDW) states by expanding the pairing amplitude in order parameters,
\begin{multline}
\label{eqn:OP_Expansion}
\Delta_{PD}(\vec{r}_{1},\vec{r}_{2})=\langle \hat{\psi}^{\dagger}_{\sigma}(\vec{r}_{1}) \hat{\psi}^{\dagger}_{-\sigma}(\vec{r}_{2}) \rangle = \\ F(\vec{r}_{1}-\vec{r}_{2}) \left[ \Delta^{\vec{Q_x}}_{P}(\vec{r})e^{i\vec{Q_x} \cdot \vec{r}}+\Delta^{-\vec{Q_x}}_{P}(\vec{r})e^{-i\vec{Q_x} \cdot \vec{r}} + \right. \\ \left.\Delta^{\vec{Q_y}}_{P}(\vec{r})e^{i\vec{Q_y} \cdot \vec{r}}+\Delta^{-\vec{Q_y}}_{P}(\vec{r})e^{-i\vec{Q_y} \cdot \vec{r}} \right] \textrm{ ,}
\end{multline}
where $\hat{\psi}^{\dagger}_{\sigma}(\vec{r}_{1})$ creates a quasi-particle of spin $\sigma$ at location $\vec{r}_{1}$ and $\vec{r}=(\vec{r}_{1}+\vec{r}_{2})/2$. $\Delta^{\vec{Q_x}}_{P}$ and $\Delta^{\vec{Q_y}}_{P}$ are PDW order parameters. They are complex scalar fields which carry momenta $\vec{Q_x}$ and $\vec{Q_y}$ running along orthogonal directions $\vec{x}$ and $\vec{y}$. 

Here we have chosen to consider a tetragonal system with quasi two-dimensional order so that the PDW wave-vectors lie in the square planes of the tetragonal lattice. We will first consider PDW with axial wave-vectors $\vec{Q_{\pm x}}$ and $\vec{Q_{\pm y}}$ that run along the two symmetry equivalent Cu-O directions in the CuO$_2$ planes.

The function $F(\vec{r}_{1}-\vec{r}_{2})$ is the form factor of the PDW. Because the axial wave-vectors break rotational symmetry means the form factors are not themselves sufficient to determine which irreducible representation of the point group the PDW transforms as. However, if $F(\vec{r}_{1}-\vec{r}_{2})$ is even under 90$^{\circ}$ rotation then the PDW can be termed to have an \emph{s}-wave form factor whereas if it is odd it has a \emph{d}-wave form factor. 

\subsection{2. Induced Orders at Superconducting Vortex}
In this section we describe how an order parameter that competes with superconductivity can be induced at superconducting vortices. Following reference \cite{Kivelson2002}, the Ginzburg-Landau free energy functional describing the competition between uniform superconductivity and another order parameter is given by
\begin{equation}
    \mathcal{F}[\Delta_{SC},\Delta_A]=\mathcal{F}_{\Delta_{SC}}[\Delta_{SC}]+\mathcal{F}_{\Delta_A}[\Delta_A]+u_{1}|\Delta_{SC}|^2|\Delta_A|^2+\textrm{...,}
\end{equation}
where $\Delta_{SC}$ is a complex scalar field representing the uniform superconducting order parameter and and $\Delta_A$ is a field representing a competing order such as PDW or CDW. For the case of competing order we focus on here $u_{1} >0$. 

The superconducting contribution to the free energy is given by its usual form
\begin{equation}
    \mathcal{F}[\Delta_{SC}]=\frac{\kappa_0}{2}\left| \left( \frac{\vec{\nabla}}{i} - \frac{2e}{c}\vec{A}\right)\Delta_{SC} \right|^2- \frac{\kappa_0}{4\xi^2}|\Delta_{SC}|^2+\frac{1}{4}|\Delta_{SC}|^4+\textrm{...}
\end{equation}
where $\vec{A}$ is the magnetic vector potential and $\xi$ is the superconducting coherence length. The contribution from the competing order is given by 
\begin{equation}
    \mathcal{F}_{\Delta_{A}}[\Delta_{A}]=\frac{\kappa_{\Delta_{A}}}{2}|\vec{\nabla}\Delta_{A}|^2+\frac{\alpha}{2}|\Delta_{A}|^2+\frac{1}{4}|\Delta_{A}|^4+... \textrm{ .}
\end{equation}

It is shown in reference \cite{Kivelson2002} that if and only if the subdominant competing order is sufficiently close in energy to the uniform superconducting phase is there a halo around superconducting vortices where the two orders coexist. Such a near degeneracy between uniform superconductivity and 8a0 PDW (and lack thereof with 8a0 CDW) has been demonstrated in numerical studies of the t-J model \cite{Corboz2014}. Moreover, the other alternative, an 8a$_0$ SDW order has not been reported in field-dependent neutron scattering.

 \subsection{3. Form Factor of $N(\vec{r},E)$ Modulations resulting from PDW}
In the main text we report the observation of a PDW through its attendant $N(\vec{r},E)$ modulations. Because $N(\vec{r},E)$ is a gauge invariant quantity, whereas the pairing-amplitude is not, the modulation wavelengths and form factors of the attendant $N(\vec{r},E)$ modulations follow directly from, but are not the same as, those present in the pairing amplitude.  

We can form a gauge invariant quantity from pairing amplitude by taking its modulus $|\Delta(\vec{r}_{1},\vec{r}_{2})|$. For simplicity let us consider a PDW of the form 
\begin{equation}
\Delta_{PD}(\vec{r}_{1},\vec{r}_{2})=D(\vec{r}_{1}-\vec{r}_{2})\Delta^{\vec{Q}}_{P}\cos(\vec{Q} \cdot \vec{r})
\end{equation}
where $D(\vec{r}_{1}-\vec{r}_{2})$ is a \emph{d}-wave form factor. Gauge invariant quantities derived from this, such as $N(\vec{r},E)$, would modulate as
\begin{align}
|\Delta_{PD}(\vec{r}_{1},\vec{r}_{2})|&=\sqrt{D^2(\vec{r}_{1}-\vec{r}_{2})|\Delta^{\vec{Q}}_{P}|^2\cos^2(\vec{Q} \cdot \vec{r})} \\
&=S(\vec{r}_{1}-\vec{r}_{2})|\Delta^{\vec{Q}}_{P}|\left[1+\cos(2\vec{Q} \cdot \vec{r})+...\right]
\end{align}
where $S(\vec{r}_{1}-\vec{r}_{2})$ is an \emph{s}-wave form factor. Thus, the $N(\vec{r},E)$ modulation resulting from a \emph{d}-wave PDW of wavevector $\vec{Q}$ will be s form factor and have wave-vector $2\vec{Q}$. 

Now consider the case relevant to the vortex halo where PDW and uniform \emph{d}-wave superconductivity coexist. The pairing amplitude can now be written
\begin{equation}
\Delta_{PD}(\vec{r}_{1},\vec{r}_{2})=D(\vec{r}_{1}-\vec{r}_{2})\Delta_{SC} + D(\vec{r}_{1}-\vec{r}_{2})\Delta^{\vec{Q}}_{P}\cos(\vec{Q} \cdot \vec{r})
\end{equation}
where $\Delta_{SC}$ is the uniform superconducting order parameter. Taking the modulus of the pairing amplitude to find the periodicities present in the $N(\vec{r},E)$ we obtain (omitting uniform components)
\begin{multline}
|\Delta_{PD}(\vec{r}_{1},\vec{r}_{2})|\propto S(\vec{r}_{1}-\vec{r}_{2})|\Delta^{\vec{Q}}_{P}|\cos(2\vec{Q} \cdot \vec{r})+\\
S(\vec{r}_{1}-\vec{r}_{2})|\Delta^{\vec{Q}}_{P}||\Delta_{SC}|\cos(\vec{Q} \cdot \vec{r})+...
\end{multline}
Thus, the $N(\vec{r},E)$ modulations attendant to \emph{d}-wave superconductivity coexisting with a \emph{d}-wave form factor PDW at wavevector $\vec{Q}$ are \emph{s}-wave form factor at wavevectors $\vec{Q}$ and $2\vec{Q}$ in agreement with the findings of the main text. If, instead the PDW had an \emph{s}-wave form factor then the attendant $N(\vec{r},E)$ modulations at $\vec{Q}$ would have had a \emph{d}-wave form factor.

 \subsection{4. Sub-Unit-Cell Resolution Field Dependent Imaging}

\begin{figure*}[htbp]
\includegraphics[width=17.2cm]{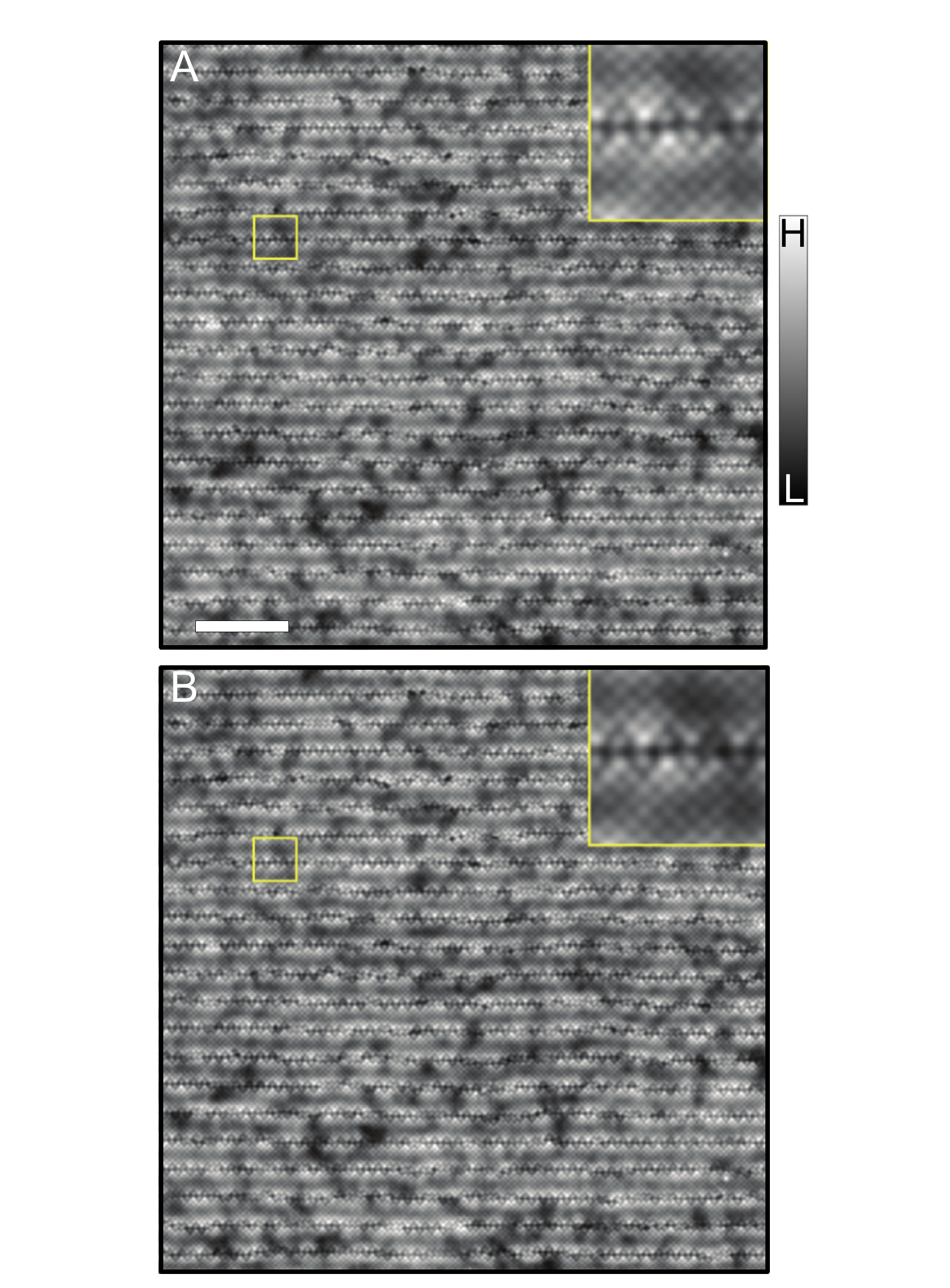}
\caption[Spatially Registered $B=8.25\textrm{T}$ and $B=0\textrm{T}$ Topographs] {\label{SI:FIG:toporeg} Spatially registered topographs of Bi$_2$Sr$_2$CaCu$_2$O$_{8+\delta}$ taken at \textbf{(A)} $B=8.25\textrm{T}$ and \textbf{(B)} $B=0\textrm{T}$. The images show a 65nmx65nm region of the sample and are registered to within 30pm. Insets show magnified images of the region indicated by the yellow squares. The scale bar is 10nm long. }
\end{figure*}

To demonstrate the high precision of spatial registration between the data sets taken at $B = 0\textrm{T}$ and $B = 8.25\textrm{T}$, we show in figure S\ref{SI:FIG:toporeg} the processed topographic images acquired simultaneously with the spectroscopic maps analyzed in the main text. These data sets were taken 2 weeks apart in the same region of the sample.  The raw data for all data sets were phase corrected using the Lawler-Fujita distortion-correction algorithm \cite{Fujita2014a}, mapping the data onto a perfectly periodic lattice free of lattice distortions due to systematic measurement effects.  A morphing scheme was then implemented to register all data sets in the same field of view (FOV) to one another with $\approx$ 30 picometer precision.  This method allows meaningful subtraction of high and low field data to detect magnetic field induced differences of the electronic structure at the sub-unit-cell scale.

\subsection{5. Energy Dependence of $\delta g(\vec{r},E)$}

\begin{figure*}[htb]
\includegraphics[width=17.2cm]{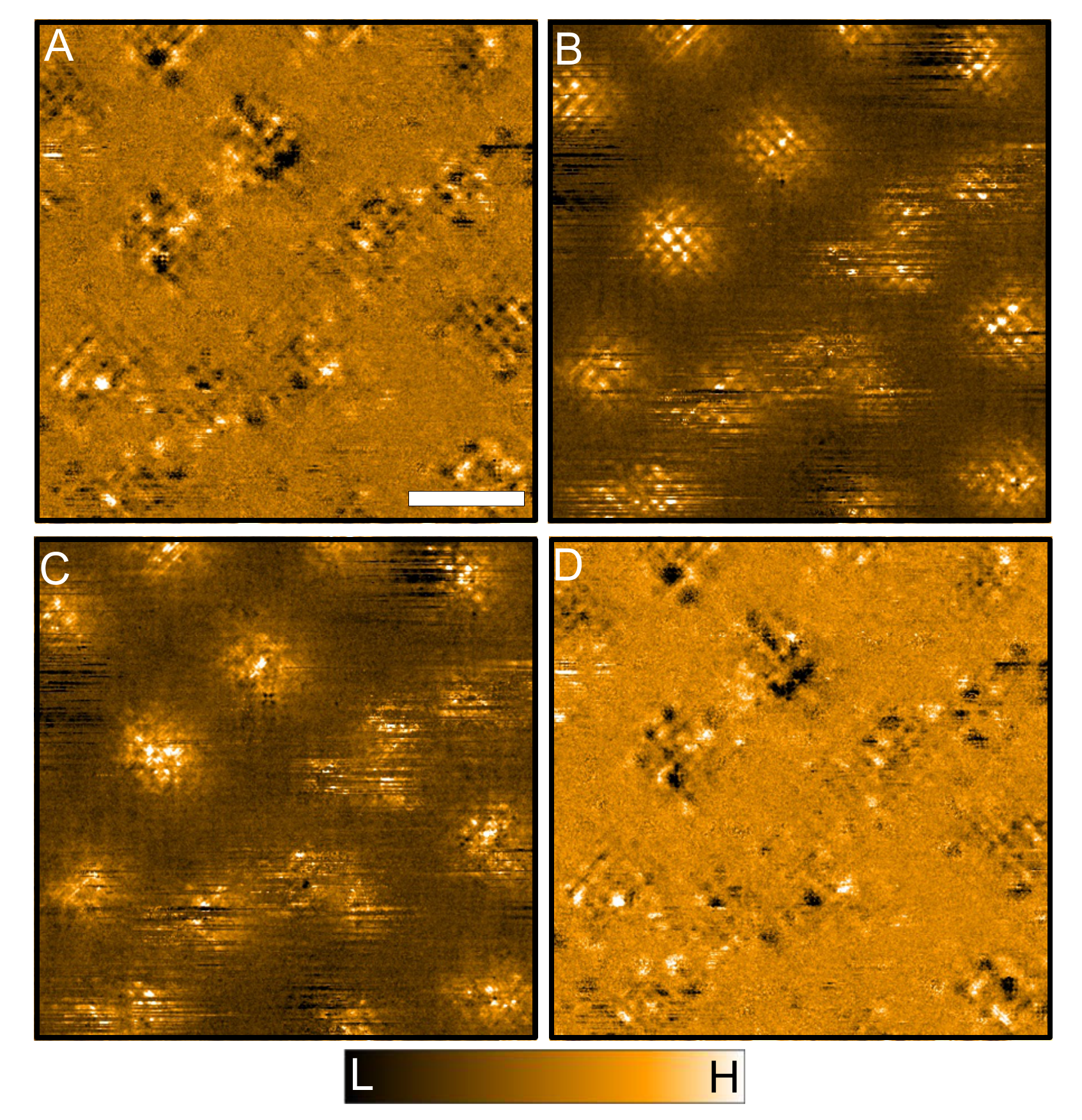}
\caption[Magnetic Field Dependent Spectroscopy] {\label{SI:FIG:greg} $\delta g(\vec{r},E)$ maps for the field of view analyzed in the main text at \textbf{(A)} $E=+30\textrm{meV}$ \textbf{(B)} $E=+10\textrm{meV}$ \textbf{(C)} $E=-10\textrm{meV}$ \textbf{(D)} $E=-30\textrm{meV}$. The scale bar is 15nm long. }
\end{figure*}

In figures S\ref{SI:FIG:greg} A-D we show $\delta g(\vec{r},E)$ for $E=+30,+10,-10,-30\textrm{meV}$ respectively. Each clearly shows magnetic field induced changes in the density of states at spatially isolated sites corresponding to superconducting vortices. 

\subsection{6. Measuring Gap Modulations}

To detect the field induced gap modulations reported in figure 5E of the main text one needs to measure the gap as a function of position for both $B=0\textrm{T}$ and $B=8.25\textrm{T}$. For each position $\vec{r}_i$ in a given differential conductance map we estimate the gap as follows: 

\begin{itemize}
    \item Find the $E_i$ that has the maximal value of $g(\vec{r}_i,E_i,B)$ for $E>0$ and denote this $E_{\Delta}$.  
    \item Fit a quadratic function to the set of the points \{$E_{\Delta-1}$,$E_{\Delta}$,$E_{\Delta+1}$\} and let the value of $E$ at which these functions are maximal be denoted $\Delta(\vec{r}_i,B)$.
\end{itemize}

The field induced gap modulations are the revealed by calculating $\widetilde{\delta\Delta}(\vec{q})=\textrm{FT}\{\Delta(\vec{r}_i,B=8.25\textrm{T})-\Delta(\vec{r}_i,B=0\textrm{T})\}$ (where FT denotes the Fourier transform) as shown in figure 5E of the main text.

\subsection{7. Bidirectional vs. Unidirectional PDW}

In the main text we show evidence that in \bscco a magnetic field induces an 8a$_0$ period PDW with approximately equal amplitude along both $\vec{Q_{x}}$ and $\vec{Q_{y}}$ when averaged over the entire field of view. If there is no long range spatial phase coherence between the PDW halos induced at each vortex, this phenomenology can arise from two scenarios. 

In the first scenario, each vortex halo contains a unidirectional PDW of the form
\begin{equation}
\Delta_{PD}(\vec{r}_{1},\vec{r}_{2})=D(\vec{r}_{1}-\vec{r}_{2}) \left[ \Delta^{\vec{Q}}_{P} e^{i\vec{Q} \cdot \vec{r}}+\Delta^{-\vec{Q}}_{P}e^{-i\vec{Q} \cdot \vec{r}} \right]
\end{equation}
where $\vec{Q}$ is either $\vec{Q_x}$ or $\vec{Q_y}$, with equal numbers of vortices choosing each of these wave-vectors. In the second scenario each vortex halo contains a bidirectional PDW of the form
\begin{multline}
    \Delta_{PD}(\vec{r}_{1},\vec{r}_{2})=D(\vec{r}_{1}-\vec{r}_{2}) \left[ \Delta_{\vec{Q_{x}}} e^{i\vec{Q_{x}} \cdot \vec{r}}+\Delta_{\vec{Q_{x}}}^{*}e^{-i\vec{Q_x} \cdot \vec{r}}+ \right. \\ \left. \Delta_{\vec{Q_{y}}} e^{i\vec{Q_{y}} \cdot \vec{r}}+\Delta_{\vec{Q_{y}}}^{*}e^{-i\vec{Q_y} \cdot \vec{r}} \right]
\end{multline}
 
To distinguish between these scenarios we must establish whether the modulations along $\vec{Q_{x}}$ and $\vec{Q_{y}}$ coexist in each vortex halo or are spatially exclusive. To this end we can calculate the function
\begin{equation}
F(\vec{r})=\frac{A_{\vec{Q_{x}}}(\vec{r})-A_{\vec{Q_{y}}}(\vec{r})}{A_{\vec{Q_{x}}}(\vec{r})+A_{\vec{Q_{y}}}(\vec{r})}
\end{equation}
where $A_{\vec{Q_{x}}}(\vec{r})$ is the local amplitude of $N(\vec{r},E)$ modulations at wavevector $\vec{Q}$, as determined using the procedure given in reference \cite{Hamidian2015}. This function measures the local imbalance in amplitude along $\vec{Q_{x}}$ and $\vec{Q_{x}}$. 

In the case of a bidirectional PDW the distribution of $F$ values in the vortex halos should be centered on $F=0$ with a standard deviation much less than 1. In the case of a unidirectional PDW which randomly picks one of two directions in each halo the distribution should have significant weight near $|F|=1$. 

\begin{figure}[htbp!]
\includegraphics[width=8cm]{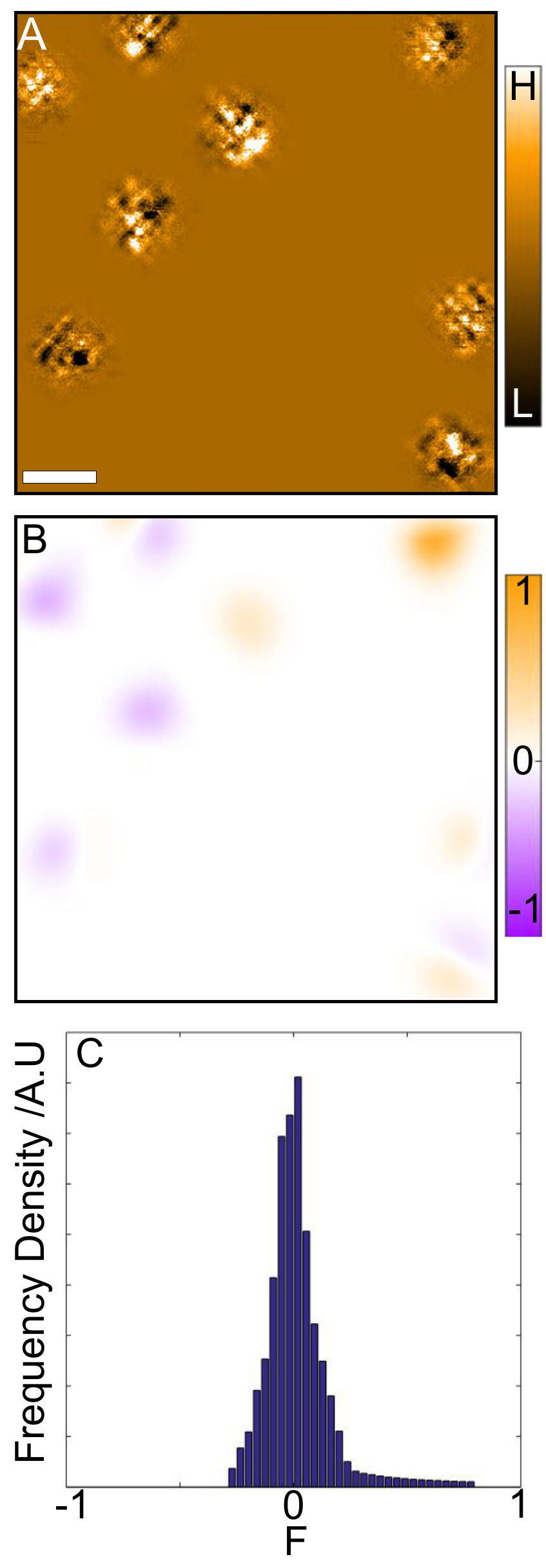}
\caption[ Bidirectional vs Unidirectional PDW] {\label{SI:FIG:dir} \textbf{(A)} $g(\vec{r},+30\textrm{meV})$ masked so as to only show a 10nm region around the center of each vortex. The scale bar is 10nm in length. \textbf{(B)} $F(\vec{r})$ derived from \textbf{A}. \textbf{(C)} Histogram of $F$ values at all pixels within 10nm of vortex center.  }
\end{figure}

In figure S3 A we show $g(\vec{r},+30\textrm{meV})$ masked so as to only show a 10nm region around the center of each vortex. In figure S3 B we show the corresponding map of $F(\vec{r})$ derived from A. While each vortex does show some amplitude imbalance in favor of one $\vec{Q_x}$ or $\vec{Q_y}$, this imbalance is small, indicating that $\vec{Q_x}$ and $\vec{Q_y}$ modulations coexist within each vortex halo. In figure S3 C we show the histogram of $F$ values from all pixels within 10nm of a vortex center. This shows distribution of $F$ values centered on $F=0$ with a standard deviation of 0.08. While it has been shown that in the presence of disorder it is difficult to distinguish unidirectional and bidirectional density wave states in $g(\vec{r},E)$ \cite{Robertson2006}, the distribution in figure in figure S3 C is most consistent with a bidirectional state.

\end{document}